\documentclass[twocolumn,showpacs,amsmath,amssymb,superscriptaddress,longbibliography]{revtex4-1}
 
\usepackage[colorlinks=true,urlcolor=blue,citecolor=blue,linkcolor=blue]{hyperref}
\usepackage{color}
\usepackage{soul}

\usepackage{amssymb,amsmath}
\usepackage{amsbsy}
\usepackage{graphicx}
\usepackage{epstopdf}
\usepackage{float}
\usepackage{pslatex}
\usepackage[usenames,dvipsnames]{xcolor}
\usepackage{lineno}
\usepackage[ansinew]{inputenc}

\newcommand{\ket}[1]{|#1\rangle}

\newcommand{\fref}[1]{Fig.~\ref{#1}}

\newcommand{\Fref}[1]{Figure~\ref{#1}}

\definecolor{blaa}{RGB}{153,153,255}
\definecolor{blaa}{RGB}{0,0,125}
\definecolor{filtered}{RGB}{153,0,51}
\definecolor{raw}{RGB}{255,177,100}
\setcitestyle{super}
\definecolor{filtered}{RGB}{175,0,0}
\definecolor{groen}{RGB}{0,150,0}
\definecolor{lil}{RGB}{160,33,160}
\definecolor{orang}{RGB}{255,60,0}

\renewcommand{\figurename}{\textbf{Fig.}}
\makeatletter
\def\fnum@figure{\figurename\nobreakspace\textbf{\thefigure}}
\def\@caption@fignum@sep{ $\boldmath $ }
\makeatother

\newcommand{\bhat}[1]{\hat{\mathbf{#1}}}
\renewcommand{\vr}{\mathbf{r}}

\newcommand\ea{{\em et al.}}
\newcommand\K{$\rm^{40}K$}
\newcommand\Rb{$\rm^{87}Rb$}
\newcommand\KRb{$^{40}$K+$^{87}$Rb}
\newcommand\AB{AB}
\newcommand\bra[1]{\left\langle#1\right|}
\begin{document}
\title{Observation of bound state self-interaction in a nano-eV atom collider}
\author{Ryan Thomas}\author{Matthew Chilcott}\affiliation{Department of Physics, QSO---Centre for Quantum Science, and Dodd-Walls Centre for Photonic and Quantum Technologies, University of Otago, Dunedin, New Zealand} \author{Eite Tiesinga}
\affiliation{Joint Quantum Institute and centre for Quantum Information and Computer Science, National Institute of Standards and Technology and University of Maryland, Gaithersburg, Maryland 20899, USA}\author{Amita B. Deb}
\author{Niels Kj{\ae}rgaard}\email{niels.kjaergaard@otago.ac.nz}
\affiliation{Department of Physics, QSO---Centre for Quantum Science, and Dodd-Walls Centre for Photonic and Quantum Technologies, University of Otago, Dunedin, New Zealand}
\date{\today}
\begin{abstract}
Quantum mechanical scattering resonances for colliding particles occur when a continuum scattering state couples to a discrete bound state between them.  The coupling also causes the bound state to interact with itself via the continuum and leads to a shift in the bound state energy, but, lacking knowledge of the bare bound state energy, measuring this self-energy via the resonance position has remained elusive. Here, we report on the direct observation of self-interaction by using a nano-eV atom collider to track the position of a magnetically-tunable Feshbach resonance through a parameter space spanned by energy and magnetic field. Our system of potassium and rubidium atoms displays a strongly non-monotonic resonance trajectory with an exceptionally large self-interaction energy arising from an interplay between the Feshbach bound state and a different, virtual bound state at a fixed energy near threshold.
\end{abstract}
\maketitle
Scattering resonances are an important feature of quantum mechanics and arise whenever asymptotically free particles are coupled to an unstable bound state of the system. While the underlying mechanism of the quasi-bound formation depends on the system, its energy and lifetime can generally be determined by measurements of the position of the scattering resonance and its width.

In nearly all systems, coupling between the asymptotically free states and the bound state affects the observable energy of the quasi-bound state.  In the non-relativistic theory, a pair of asymptotically free particles, A and B, colliding at a kinetic energy $E$ have both background and resonant contributions to their scattering amplitude $f(E,\Omega)$ \cite{Joachain1987,Svec2001}
%\begin{align}
%    \sigma_{\rm tot}(E,B)=& \frac{4\pi}{k} {\rm Im}[f(E,B)]\\
%{=}& 4\pi a_\text{bg}^2\left[1+\frac{2(E-E_\text{res})\gamma+\gamma^2}{(E-E_\text{res})^2+(\Gamma_\text{tot}/2)^2}\right ]\\ &+ \frac{2\pi}{k}\frac{a_\text{bg}\gamma\Gamma^\text{inel}}{(E-E_\text{res})^2+(\Gamma_\text{tot}/2)^2},\label{eq:scat_xsec}
%\end{align}
\begin{align}
f(E,\Omega) =& \underbrace{\frac{\left(e^{2i\delta_{\rm bg}(E)}-1\right)}{2ip/\hbar}+f_{\ell>0}(E,\Omega)}_{f_{\rm bg}(E,\Omega)}\nonumber\\ &+\underbrace{\left(-\frac{e^{2i\delta_{\rm bg}(E)}}{2p/\hbar}\frac{\Gamma(E)}{E-E_{\rm AB}(E)+i\Gamma(E)/2}\right)}_{f_{\rm res}(E)},
\label{eq:ScatteringAmp}
\end{align}
%\begin{widetext}
%\begin{equation}
%f(E,\Omega) = \underbrace{\frac{\left(e^{2i\delta_{\rm bg}(E)}-1\right)}{2ip/\hbar}+f_{\ell>0}(E,\Omega)}_{f_{\rm bg}(E,\Omega)}+\underbrace{\left(-\frac{e^{2i\delta_{\rm bg}(E)}}{2p/\hbar}\frac{\Gamma(E)}{E-E_{\rm AB}(E)+i\Gamma(E)/2}\right)}_{f_{\rm res}(E)},
%\label{eq:ScatteringAmp}
%\end{equation}
%\end{widetext}
where $p=\sqrt{2mE}$ is the magnitude of the relative momentum of the two particles with $m$ the reduced mass, $\delta_{\rm bg}(E)$ is the s-wave background scattering phase shift, $\Gamma(E)$ is an energy-dependent resonance width, and $E_{\textrm{AB}}(E)$ is the energy of the quasi-bound state \AB{}.
\begin{figure}[tb!]
\centering
\includegraphics[width=\columnwidth]{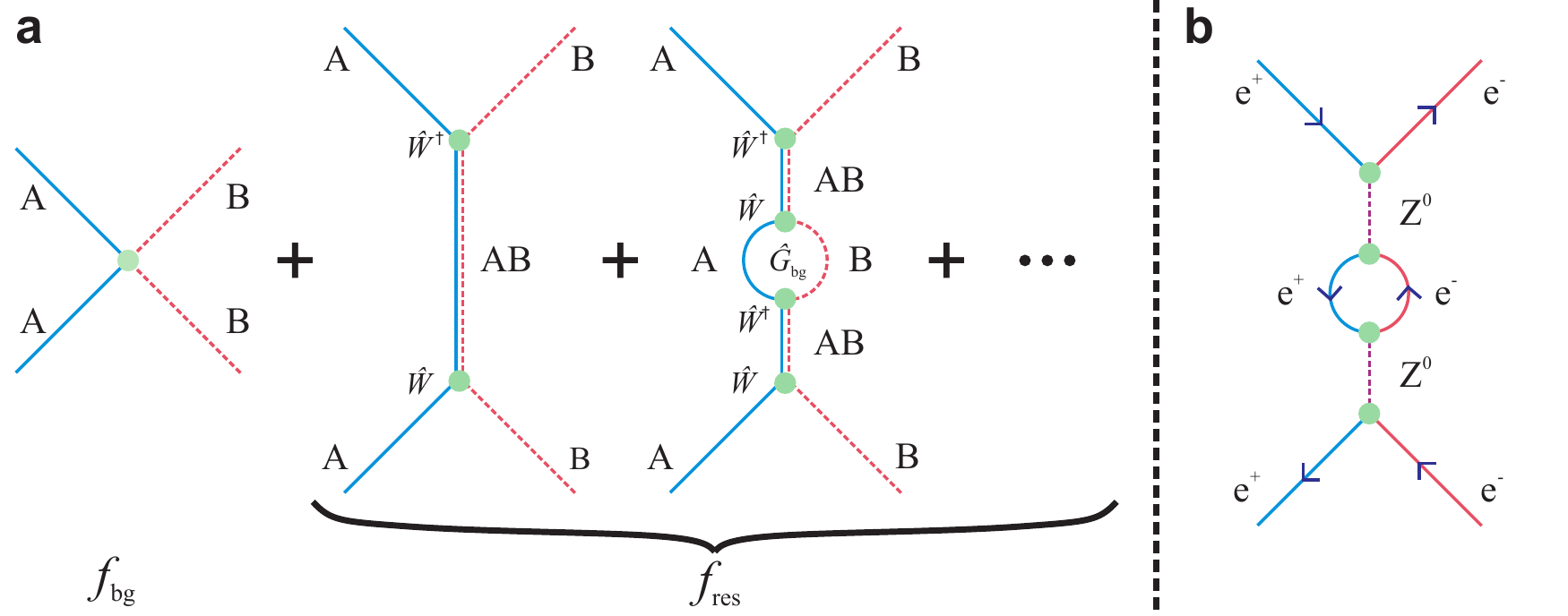}
\caption{{Feynman diagrams for resonant collisions.}  \textbf{a} The total scattering amplitude $f(E)$ for asymptotically free particles A and B is the sum of the background contribution $f_{\rm bg}$ and the resonant contribution $f_{\rm res}$.  The lowest order contribution to $f_{\rm res}$ has A and B forming the quasi-bound state \AB{} which then decays back to the individual particles.  The Eq.next order adds a self-energy correction to the bound state energy.  \AB{} decays into particles A and B which reform into the quasi-bound state \AB{}.  \textbf{b}  Analogous self-energy correction for the mass of the Z$^0$ boson in the elastic scattering of electrons and positrons.}
\label{fg:Feynman}
\end{figure}
Higher angular momentum scattering amplitudes are included in the direction-dependent term $f_{\ell>0}(E,\Omega)$ where $\Omega=(\theta,\varphi)$ specifies the direction of the outgoing particles in spherical polar coordinates.  \Fref{fg:Feynman}a represents Eq.~\eqref{eq:ScatteringAmp} diagrammatically.   The background scattering amplitude, $f_{\rm bg}(E,\Omega)$, is added to the resonant scattering amplitude, $f_{\rm res}(E)$, which is an infinite sum of contributions from different order processes that start and end with asymptotically free particles A and B.  When only the leading order, zero-loop term is present then $E_{\textrm{AB}}(E)=E_{0}$ is the bound state energy in the absence of coupling.  However, if we include all loop orders, then the quasi-bound state's energy becomes \cite{Friedrich2016}
\begin{equation}
E_{\textrm{AB}}(E) = E_{0} +\underbrace{ \bra{\textrm{AB}}\hat{W}\hat{G}_{\rm bg}(E)\hat{W}^\dag\ket{\textrm{AB}}}_{\delta E(E)}.
\label{eq:BoundEnergy}
\end{equation}

The interpretation of the energy shift $\delta{E}$ is that the bare bound state $\ket{\textrm{AB}}$ dissociates into free particles by the action of coupling operator $\hat{W}^\dag$, the individual particles propagate according to the background Green's operator $\hat{G}_{\rm bg}(E)$, after which they associate back into the bound state by $\hat{W}$.  This process is similar to self-energy effects that alter the mass of force-carrying particles such as the Z$^0$ boson (\fref{fg:Feynman}b)  in relativistic electron-positron interactions\cite{Novikov1999}.  In all cases, when the bare energy of the intermediate state is fixed we can only measure the re-normalized, ``dressed'' bound state energy $E_{\textrm{AB}}$ : it is impossible to observe the actual self-energy correction.

In this study, we use the tunability of a magnetic Feshbach resonance\cite{Feshbach1,Feshbach2,Fano,FeshbachReview} in atomic collisions to directly observe self-energy corrections to the quasi-bound state energy.  Through the application of a magnetic field $B$, the energy of the bare (uncoupled) bound state can be varied according to $E_{0} = \delta\mu (B-B_c)$, where $\delta\mu$ is the difference in magnetic moments between the free atoms and the molecular state AB, and $B_c$ is the field at which the bound state energy is equal to the background channel's threshold energy.  This tunability has made Feshbach resonances pivotal in modern atomic physics, as one can tailor the interactions for ultracold and quantum degenerate gases \cite{InouyeAndrewsStengerEtAl1998}, but it also means that, with few exceptions\cite{Durr2004,Volz2005,Gensemer2012,Genkina2016,Horvath2017}, the study of Feshbach resonances in ultracold gases has invariably been conducted on trapped samples, where the atomic collision energy is defined by the sample temperature.  As temperatures are sufficiently low for collisions to remain in the threshold regime, such experiments cannot provide insight into self-energy effects.  Instead, one must probe the resonance over a range of both collision energy and magnetic field.  Substituting $E_0=\delta\mu(B-B_c)$ into Eq.~\eqref{eq:BoundEnergy}, and the result into Eq.~\eqref{eq:ScatteringAmp}, the resonant scattering amplitude becomes
\begin{equation}
f_{\rm res}(E,B) = \frac{e^{2i\delta_{\rm bg}(E)}}{2p/\hbar}\frac{i\Gamma_B(E)}{B-B_{\rm res}(E)-i\Gamma_B(E)/2},
\label{eq:ResonantPhase2}
\end{equation}
where $\Gamma_B(E)=\Gamma(E)/\delta\mu$ and $B_{\rm res}(E) = B_c+\left[E-\delta E(E)\right]/\delta\mu$ is the resonant magnetic field.  By measuring $B_{\rm res}(E)$ at sufficiently high energies, we can infer both $\delta\mu$ and $B_c$, since $\delta E (E) \stackrel{E\rightarrow\infty}{\rightarrow} 0$\cite{Julienne1989}, and hence determine the self-energy of the quasi-bound state.
\section*{Results}
\noindent\textbf{System under study.}
We explore the collision energy and magnetic field dependence of an s-wave Feshbach resonance in \K{}+\Rb{} using an optical collider\cite{Rakonjac2012a,Thomas2016,Horvath2017}.  This resonance occurs between \Rb{} atoms in the $\ket{F=1,m_F=1}$ state and \K{} atoms in the $\ket{\frac{9}{2},-\frac{9}{2}}$ state near 546 G ($10^4~\textrm{G}=1~\textrm{T}$)\cite{Inouye2004,Ospelkaus2006,Ferlaino2006,Klempt2008}.  As these states are the absolute ground states of \K{} and \Rb{} all collisions are elastic, and the cross section at a collision energy $E$ near the resonance can be written as\cite{Friedrich2016,FeshbachReview}
\begin{align}
\sigma(E,B) =& \int \left|f(E,\Omega)\right|^2d\Omega\nonumber\\
=& \frac{4\pi\hbar^2}{mE}\sin^2\left[\delta_{\rm bg}(E)+\tan^{-1}\left(\frac{\Gamma_B(E)/2}{B-B_{\rm res}(E)}\right)\right] \nonumber\\&+ \sigma_{\ell>0}(E),
\label{eq:CrossSec}
\end{align}
where the first term is a Beutler-Fano profile\cite{Friedrich2016,Fano,Beutler1935} as a function of $B$ for a given energy $E$, and $\sigma_{\ell>0}(E)$ accounts for the non-zero contribution to the cross section from higher ($\ell>0$) partial waves when $E$ is larger than the Wigner threshold energy, which is $E/k \approx 100$ $\mu$K for \KRb{}, where $k$ is the Boltzmann constant. The magnetic field dependence of the cross section is experimentally determined through measurements of the fraction of atoms scattered in a collision of two cold atomic clouds
\begin{equation}
S(E,B)=\frac{\alpha(E)\sigma(E,B)}{1+\alpha(E)\sigma(E,B)},
\label{eq:ScattFrac}
\end{equation}
where $\alpha(E)$ is scaling factor determined by the peak densities of the clouds and their transverse spatial overlap at the time of collision (see Supplementary Note 1).  By measuring $S(E,B)$ as a function of $B$ for fixed $E$, we determine $\delta_{\rm bg}(E)$, $\Gamma_B(E)$, and $B_{\rm res}(E)$.

%Furthermore, a measurement of $S$ for a single energy determines only the relative sign between $\delta_{\rm bg}$ and $\Gamma_B$; by measuring at a number of sufficiently high energies, we fix the sign of $\delta\mu$ and hence the absolute signs of $\delta_{\rm bg}$ and $\Gamma_B$.  This interferometric determination of the resonance parameters is valid for any isolated resonance with negligible inelastic scattering.
\begin{figure}[tb!]
\centering
\includegraphics[width=\columnwidth]{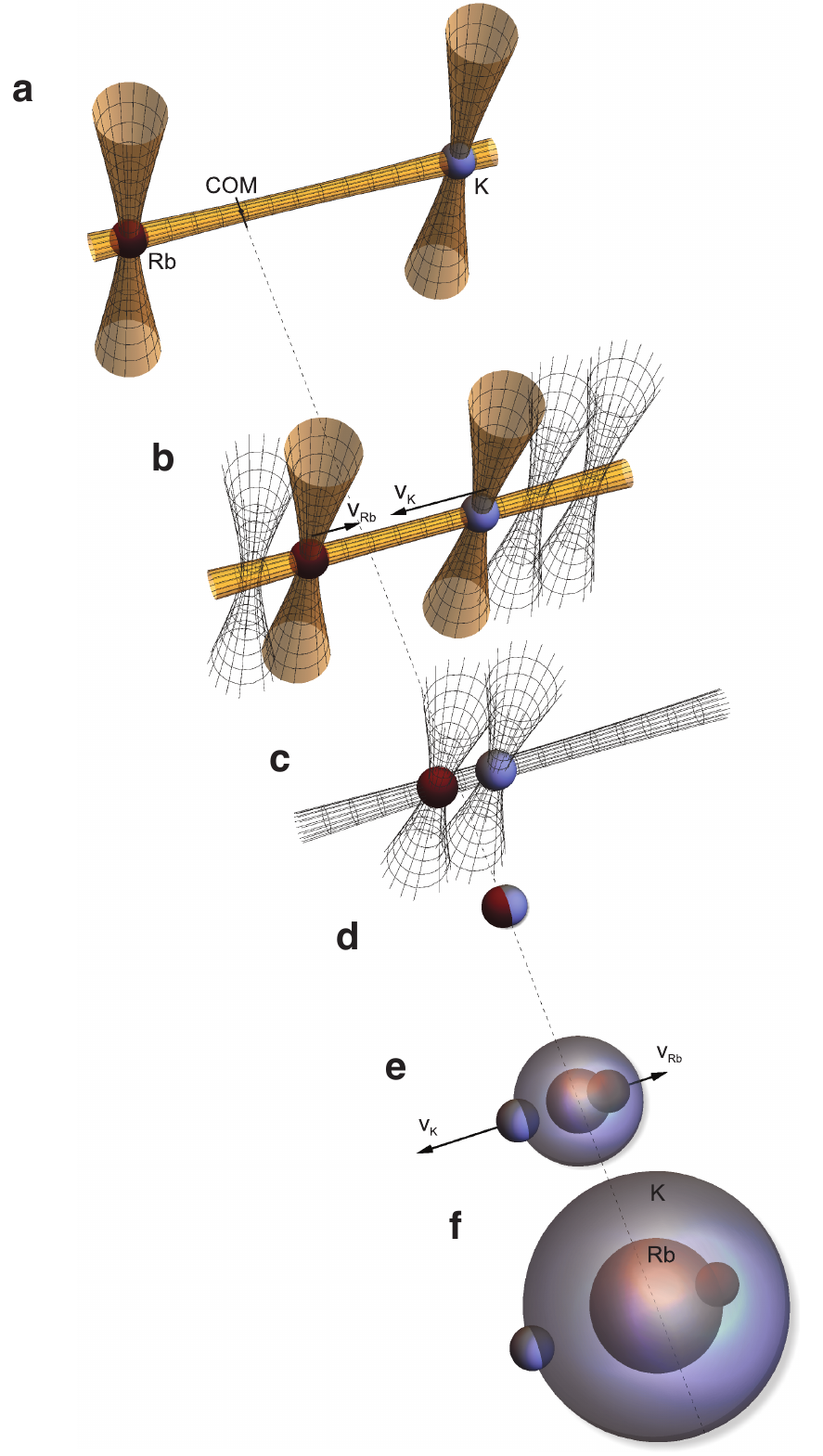}
\caption{{Optical collider procedure.} \textbf{a} \K{} (blue, right) and \Rb{} (red, left) atoms, prepared in the desired internal quantum states, are held in two crossed optical dipole traps separated by approximately $3$ mm; COM indicates the center-of-mass for pairs of K and Rb atoms.  \textbf{b}~The two traps are accelerated towards each other, keeping $m_{\rm Rb}v_{\rm Rb}=-m_{\rm K}v_{\rm K}$, so that pairs of of K and Rb atoms have zero total momentum on average, and their COM is at rest in the laboratory frame.  {\textbf{c}~When} the wells are separated by $\approx 60$ $\mu$m, the optical traps are switched off.  \textbf{d}  The two clouds collide in free space.  \textbf{(e,f}, The K and Rb collision halos expand at different rates.  We image the K halo at the time represented in (\textbf{e}), and then wait to image Rb until its halo has expanded to the equivalent size (\textbf{f}).}
\label{fg:setu}
\end{figure}
%\FloatBarrier
\noindent\textbf{Experimental procedure.}
\begin{figure*}[!tb]
\centering
\includegraphics[width=0.95\textwidth]{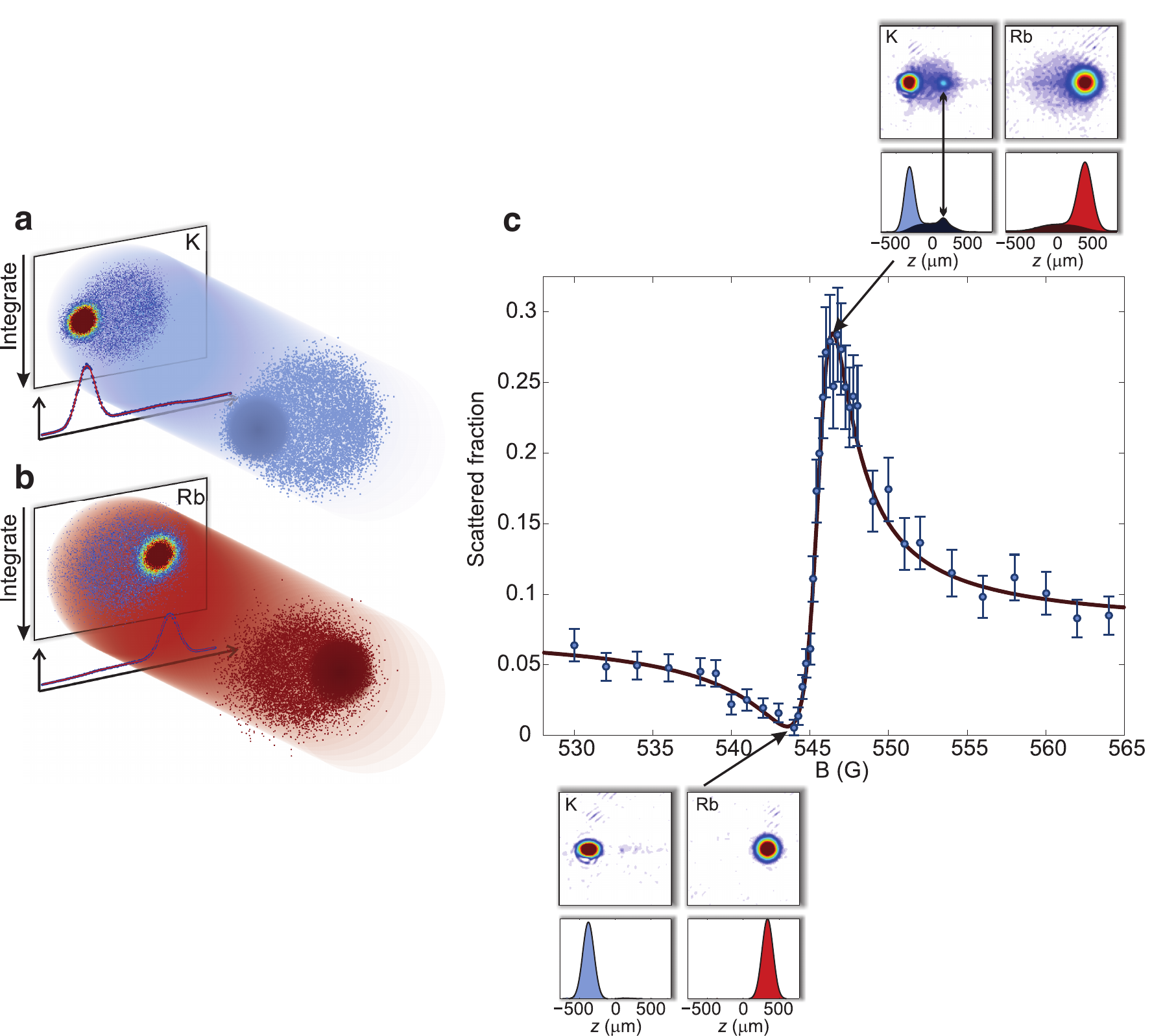}
\caption{{Imaging and analysis of \KRb{} collisions.}  \textbf{a,b} 3D collision halos are projected onto a 2D plane by absorption imaging.  We integrate the projected images in the vertical direction to obtain line densities.  Line densities are fitted to extract the scattered fraction.  \textbf{c}~Measured scattered fraction (circles) as a function of magnetic field for $E/k=52(1)$ $\mu$K; error bars indicate one standard deviation uncertainties.  Solid line is a fit to Eq.~\eqref{eq:CrossSec} and Eq.~Eq.\eqref{eq:ScattFrac}.  Insets show absorption images and density profiles of K (left) and Rb (right) at the indicated magnetic fields.  Light blue and red coloured areas in density profiles indicate unscattered atoms and dark red and blue areas scattered atoms.  Double-headed arrow indicates an additional \K{} feature (see text and Supplementary Fig. 1).}
\label{fg:Example}
\end{figure*}
We begin our experiment by loading an ultracold mixture of \K{} atoms in the $\ket{\frac{9}{2},\frac{9}{2}}$ state and \Rb{} atoms in the $\ket{2,2}$ state into a double-well far off-resonant optical dipole trap.  The double well is formed by the intersection of a horizontal laser beam with two vertical laser beams generated by rapidly switching between two frequency pairs that drive a two-axis acousto-optic deflector\cite{Roberts2014,Thomas2016}.  The initial well separation is 80 $\mu$m, or approximately twice the vertical laser beam waist.  We transfer Rb atoms to the $\ket{1,1}$ state and separate the double well in the presence of a strong magnetic field gradient which preferentially pushes Rb atoms into one well and K atoms into the other.  The two wells are moved to final positions $z_{\rm K}=2.0$ mm for K and $z_{\rm Rb}=-(m_{\rm K}/m_{\rm Rb}) z_{\rm K}=-0.92$ mm for Rb, where $m_i$ are the masses of \K{} and \Rb{}, respectively.  We prepare the internal states of each species of atoms using microwave and radio-frequency transitions between hyperfine states at a magnetic field of approximately $9$ G, and we purge the wells of unwanted states and species using resonant light pulses.  At the end of the state preparation, we have $3.0(3)\times 10^5$ atoms of each species in their respective wells at temperatures of 1.1(1) $\mu$K for K and $0.8(1)$ $\mu$K for Rb, as determined by time-of-flight absorption images.  Throughout this work, numbers in parentheses correspond to the one standard deviation uncertainty combining statistical and systematic contributions.  With the atoms in the desired internal quantum states, we use a dedicated, water-cooled coil pair in the Helmholtz configuration to create a stable and homogeneous magnetic field around the 546 G Feshbach resonance.  The current in these coils is regulated to a fractional stability of $<10^{-5}$, and the magnetic field is calibrated using Rabi spectroscopy between the \Rb{} $\ket{2,0}$ and $\ket{1,0}$ states to an accuracy of 5 mG.

\Fref{fg:setu} shows the operation (see also  \href{http://www.physics.otago.ac.nz/staff_files/nk/files/MovieS1.mp4}{Supplementary Movie 1}) of our optical collider.  From their initial positions in \fref{fg:setu}a, the atomic clouds of Rb and K are accelerated towards each other along the horizontal guide beam by steering the vertical beams of the crossed traps as illustrated in \fref{fg:setu}b. The traps are moved at velocities $v_\text{Rb}$ and $v_\text{K}$ such that $v_\text{Rb}/v_\text{K}=-m_\text{K}/m_\text{Rb}$. This keeps the center-of-mass (COM) of the collisional partners at rest in the laboratory frame and ensures that expanding spheres of scattered particles are centered around a common point. Just before the collision we turn off all laser beams (\fref{fg:setu}c), so that the atoms collide in free space (\fref{fg:setu}d).  After the collision, we switch the magnetic field off (see ``Methods'') and wait until the K cloud has moved 350 $\mu$m from the collision point. At this time a halo of scattered particles will have formed (\fref{fg:setu}e), and we record the distribution of K atoms using absorption imaging along the $x$-axis as in Fig.~\ref{fg:Example}a.  We then wait until the Rb atoms have also moved 350 $\mu$m from the collision point (\fref{fg:setu}f) and acquire an image of the Rb atoms (\fref{fg:Example}b); a frame-transfer CCD allows us to acquire images of both species of atoms in rapid succession.  We determine the collision energy by acquiring two pairs of images with a time-of-flight difference of 5 ms, measuring the distance travelled by the unscattered atoms in that time, and then calculating the kinetic energy.  The uncertainty in the determination of the collision energy from both systematic and statistical sources is 2\% of the value.

\noindent\textbf{Measurements.}
Using our optical collider, we measure the scattered fraction of atoms $S$ from absorption images (see ``Methods'') as a function of magnetic field with the \K{}+\Rb{} system prepared in states $\ket{\frac{9}{2},-\frac{9}{2}}$ (\K{}) and $\ket{1,1}$ (\Rb{}) for collision energies $E/k=10$ $\mu$K to 300 $\mu$K. \Fref{fg:Example}c presents example data acquired at collision energy $E/k=52(1)~\mu$K. While this energy, equivalent to 4.5~neV, is far below that of conventional particle colliders, it remains two orders of magnitudes higher than the typical energy scales for studies of Feshbach resonances in trapped ultracold atomic gases as set by the sample temperature. The fraction $S$ has a pronounced Beutler-Fano lineshape associated with an interspecies Feshbach resonance centered on a magnetic field of approximately $546$~G.  The insets of \fref{fg:Example}c show post-collision absorption images of \K{} and \Rb{} for the field values where $\sigma$ (and hence $S$) attains its maximum and minimum.  At the minimum, no discernible scattering halo is visible as a result of destructive quantum interference, where the background phase shift $\delta_{\rm bg}(E)$ of Eq.~(\ref{eq:CrossSec}) is cancelled by the resonant contribution; only outgoing clouds of unscattered particles are visible in the absorption images. For the maximum, where the interference is constructive, as well as away from resonance, a halo of isotropically scattered particles emerges. The \K{} images also reveals an additional feature located at the position of the outgoing \Rb{} cloud caused by multiple scattering (see also Supplementary Fig.~1 which shows the result of a numerical simulation elucidating the effect of multiple scattering).

\begin{figure}[!htbp]
\centering
\includegraphics[width=0.6\columnwidth]{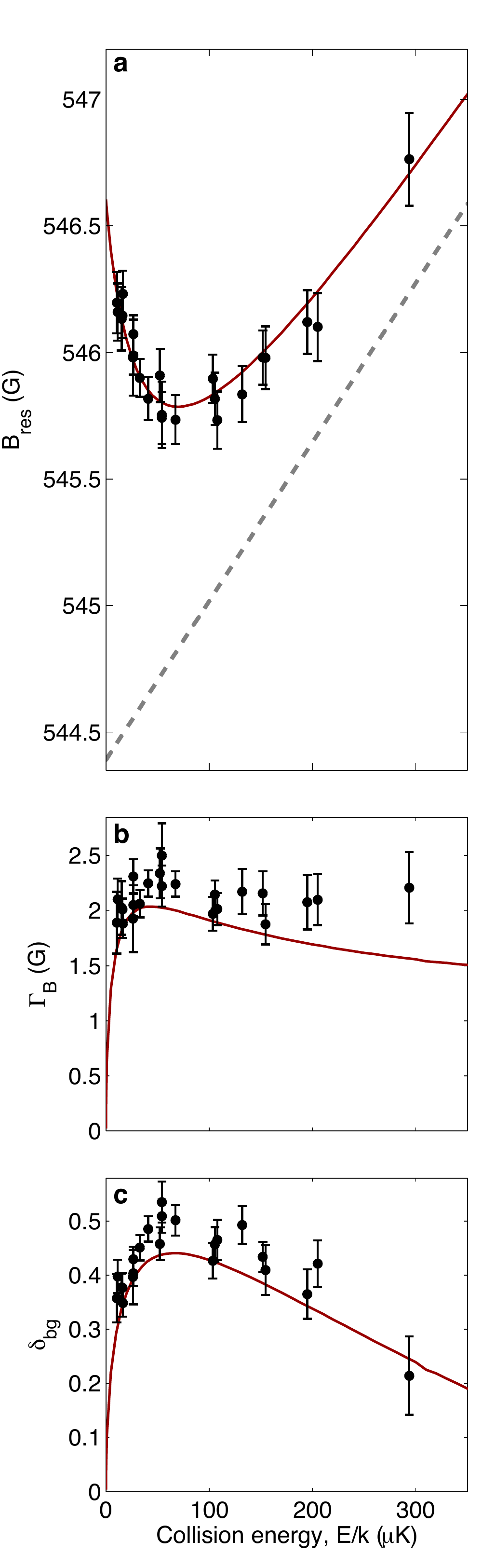}
\caption{Parameters describing the \KRb{} Feshbach resonance.  \textbf{(a-c} Resonance position $B_{\rm res}$, width $\Gamma_B$, and background phase shift $\delta_{\rm bg}$ as a function of collision energy $E$ for the resonance near 546 G.  Solid lines are predictions from a coupled-channels model.  Circles are measurements using the collider, and error bars are the one standard deviation uncertainty obtained from the fits to Eq.~\eqref{eq:CrossSec} and Eq.~\eqref{eq:ScattFrac}.  The dashed grey line in \textbf{(a)} indicates the resonance position in the absence of self-energy effects extrapolated from the asymptotic behaviour of the fitted model for $B_{\rm res}(E)$.}
\label{fg:Results1}
\end{figure}

\noindent\textbf{Data analysis.}
We fit Eq.~\eqref{eq:CrossSec} and Eq.~\eqref{eq:ScattFrac} to our data allowing $\alpha$, $\sigma_{\ell>0}$, $\delta_{\rm bg}$, $\Gamma_B$, and $B_{\rm res}$ to vary as functions of energy.  \Fref{fg:Example}c shows the fitted curve for the $E/k=52(1)~\mu$K case, which is typical across our entire energy range. Figures \ref{fg:Results1}a-c show our fitted values for $B_{\rm res}$, $\Gamma_B$, and $\delta_{\rm bg}$ as functions of energy along with predictions from a coupled-channels model based on published interaction potentials\cite{Pashov2007}.
\section*{Discussion}
A particularly striking feature of our measurements is that the resonance position $B_{\rm res}$ is not a monotonic function of $E$ (Fig.~\ref{fg:Results1}a).  From zero energy, $B_{\rm res}(E)$ decreases to a minimum value near $E/k=75$ $\mu$K before approaching a linear asymptote.  This curvature is solely the result of the self-interaction energy $\delta E(E)$ due to higher-order resonant scattering processes such as those shown in \fref{fg:Feynman}a.  Without these processes, the resonance position would follow the linear curve $B_c+E/\delta\mu$ in \fref{fg:Results1}a.  The gap between the two curves, therefore, is a direct measure of the self-energy $\delta E(E)/\delta\mu$.  The marked non-monotonic behaviour, as opposed to only being non-linear, is due to the negative background scattering length $a_{\rm bg}=-\hbar\lim\limits_{E\rightarrow 0} \tan\delta_{\rm bg}(E)/p$ which is associated with a ``virtual'' bound state just below threshold in the energetically open channel\cite{Marcelis2004,Bortolotti2008}, and whose effect on the open channel scattering can be seen in Fig.~\ref{fg:Results1}b and Fig.~\ref{fg:Results1}c as the similarly non-monotonic behaviours of the background phase shift and resonance width, respectively.  This is in contrast to systems with a positive background scattering length, where both the background phase shift and resonance width scale as $E^{1/2}$ for energies up to the van der Waals energy ($E_{\rm vdW}/k \approx 600$ $\mu$K for \KRb{}).
%The interplay between the virtual bound state near threshold and the quasi-bound state responsible for the Feshbach resonance leads to a large, negative self-energy that produces both the curvature and minimum in the resonance position as a function of energy. This feature is distinctly different from the case of a positive scattering length, where the trajectory of the resonance position is dominated by a Landau-Zener-like avoided crossing just below threshold, and is therefore always strictly monotonic in energy\cite{Marcelis2004}.
While measurements of molecular binding energies also probe self-energies\cite{Klempt2008,Deuretzbacher2008}, we note that this is the first direct observation of such self-interaction effects in Feshbach resonances for continuum states as previous experiments worked with resonances with widths that were too small to generate appreciable curvature\cite{Gensemer2012,Horvath2017}.

Although we have focused on one particular resonance, our technique is more general and can be applied to both isolated and overlapping Feshbach resonances.  By performing our measurements at fixed collision energy, we isolate the functional dependence of both the background and resonant scattering phases and eliminate the need for \textit{a priori} knowledge about how these parameters vary with energy.  Elastically dominated resonances can be described by a simple Beutler-Fano function, and even resonances with significant inelastic scattering have analytic descriptions as functions of magnetic field at fixed energy\cite{Hutson2007}.  The ability to study resonances with significant inelastic widths is a marked improvement over molecular-dissociation experiments, where a stable molecular state is a required starting point\cite{Durr2004,Volz2005}.  Furthermore, by measuring the number of scattered atoms we avoid the need for an external interferometer to measure the scattering phase\cite{Gensemer2012}.  Indeed, the number of scattered atoms can be regarded as the output of an ``internal'' interferometer, where the path length for one arm at a given energy is fixed and yields the background phase, and the path length of the other, resonant, arm can be varied with a magnetic field.  Given the method's generality and the relatively high energies that we can access with our collider, we expect that this work can be extended to study Feshbach resonances in higher partial waves\cite{PhysRevLett.119.203402,Yao2017}, especially in regards to self-energy effects in the interference between Feshbach and shape resonances\cite{Volz2005}, or to  study multiple scattering effects in jet production in unitary Fermi gases\cite{Pan1994}.
\vspace{4mm}
\section*{Methods}
\small\noindent\textbf{Counting of scattered atoms.}
We count atoms by acquiring absorption images of both the rubidium and potassium clouds and converting those to density plots.  The frequency of our probe laser system is not sufficiently tunable to image clouds at magnetic fields of $>50$ G, so we must turn off the Feshbach magnetic field prior to image acquisition.  After the two atom clouds collide, we wait until the initial atom clouds are separated by 160 $\mu$m before we switch off the Feshbach magnetic field, which ensures that the magnetic field is present during the entire collision.  We then wait until each cloud has travelled 350 $\mu$m from the collision point before we acquire an absorption image of the atoms; due to the differences in the speed of the Rb and K clouds, we take these images at two different times.  Depending on the collision energy, the delay  of K image acquisition with respect to turning off the Feshbach field coil varies between $0.5$ ms to 4 ms.  The large electrical currents ($>100$ A) flowing in the coil that need to be switched off, combined with metal components surrounding the atoms, mean that eddy currents, and hence stray magnetic fields, are present during imaging.  As a result, we image in an unknown magnetic field, and while we optimize the probe laser frequency for each collision energy, the absolute number of atoms that we extract from our absorption images is not directly comparable between experiments conducted at different energies.  Therefore, we calculate the fraction of atoms scattered by the collision and compare that to theoretical predictions instead of the total number of scattered atoms, as it eliminates a free parameter from our fitted model.

\vspace{4mm}
\small\noindent\textbf{Determination of scattered fraction.}
We measure the scattered fraction by integrating the absorption images along an axis perpendicular to the collision axis $z$ to produce line density distributions $n_i(z)$, where $i$ can be either K or Rb.  Each $n_i(z)$ can be expressed as the sum of three terms $n_i(z) = N_i^u P_i^u(z) + N_i^s P_i^s(z) + N_i^m P_i^m(z)$, where each probability density $P_i^x(z)$ is normalized to unity.  The distribution $P_i^u(z)$ represents the distribution of unscattered atoms and is a Gaussian distribution.  $P_i^s(z)$ describes scattered atoms whose distribution is proportional to the differential cross-section. In computing this, we neglect the initial spatial distribution and the distribution of total momentum -- both of which only lead to negligible blurring of the final pattern.  Finally, $P_i^m(z)$ accounts for multiply scattered atoms whose distribution cannot be described using the differential cross-section.  These multiply scattered atoms are mostly K atoms due to the higher density of the Rb cloud, and the potassium atoms have a Gaussian distribution centered at the position of the Rb cloud as shown in Supplementary Fig.~1. We fit $n_i(z)$ to our integrated densities to extract the number of unscattered $N_i^u$, scattered $N_i^s$, and multiply scattered $N_i^m$ atoms. The scattered fraction $S$ is then
\[
S=\frac{\sum\limits_i(N_i^s+N_i^m)}{\sum\limits_i (N_i^u+N_i^s+N_i^m)}.
\]

\vspace{4mm}
\noindent\textbf{Data availability.} The data that support the findings of this study are available
from the corresponding author on reasonable request.
%\bibliography{KRb_bib,fano2,naturefbr}
%\bibliographystyle{trybib}

\section*{Acknowledgments}
This work was supported by the Marsden Fund of New Zealand (Contract No. UOO1121).

\section*{Author Contributions}
\noindent N.K. conceived the project. R.T. performed experiments with support from A.B.D. R.T. implemented the magnetic field servo with assistance from M.C. R.T analysed the data and conducted coupled channels calculations. E.T. provided theory. R.T. and N.K. prepared the manuscript with input and comments from all authors. N.K. supervised the project.
\subsection*{Additional Information}
 \noindent Supplementary Information accompanies this paper at
\href{http://www.physics.otago.ac.nz/staff_files/nk/files/krbsupp.pdf}{www.physics.otago.ac.nz/staff\_files/nk/files/krbsupp.pdf}
%\href{http://www.physics.otago.ac.nz/staff_files/nk/files/MovieS1.mp4}{www.physics.otago.ac.nz/staff_files/nk/files/}

\vspace{4mm}
\noindent{\bf Competing financial interests:} The authors declare no competing financial interests.
\end{document}